\documentclass[%
 reprint,
 amsmath,amssymb,
pra,
]{revtex4-2}
\usepackage{xcolor}
\usepackage{graphicx}
\usepackage{dcolumn}
\usepackage{bm}

\begin{document}

\preprint{APS/123-QED}

\title{Passive decoy-state quantum secure direct communication with heralded single-photon source }
\author{Jia-Wei Ying,$^{1,2}$ Peng Zhao,$^{1,2}$ Wei Zhong,$^{2}$ Ming-Ming Du,$^{1}$, Xi-Yun Li,$^{3}$  Shu-Ting Shen,$^{1}$ An-Lei Zhang,$^{3}$ Lan Zhou,$^{3}$}\email{zhoul@njupt.edu.cn}
\author{Yu-Bo Sheng$^{1,2}$}
 \email{shengyb@njupt.edu.cn}
\affiliation{%
 $^1$College of Electronic and Optical Engineering and College of Flexible Electronics (Future Technology), Nanjing
 University of Posts and Telecommunications, Nanjing, 210023, China\\
 $^2$Institute of Quantum Information and Technology, Nanjing University of Posts and Telecommunications, Nanjing, 210003, China\\
 $^3$College of Science, Nanjing University of Posts and Telecommunications, Nanjing, 210023, China\\
}%

\date{\today}

\begin{abstract}
Quantum secure direct communications (QSDC) can directly transmit secret messages through a quantum channel without keys. The imperfect photon source is a major obstacle for QSDC's practical implementation. The unwanted vacuum state and multiphoton components emitted from imperfect photon source largely reduce QSDC's secrecy message capacity and even threaten its security. In the paper, we propose a high-efficient passive decoy-state QSDC protocol with the heralded single-photon source (HSPS). We adopt a spontaneous parametric down-conversion source to emit entangled photon pairs in two spatial modes. By detecting the photons in one of the two correlated spatial modes, we can infer the photon-number distribution of the other spatial mode. Meanwhile, our protocol allows a simple passive preparation of the signal states and decoy state.
   The HSPS can effectively reduce the probability of vacuum state and increase QSDC's secrecy message capacity. Meanwhile, the passive decoy-state method can simplify the experimental operations and enhance QSDC's robustness against the third-party side-channel attacks. Under the communication distance of 10 km, the secrecy message capacity of our QSDC protocol can achieve 81.85 times with average photon number of 0.1 and 12.79 times with average photon number of 0.01 of that in the original single-photon-based QSDC protocol without the HSPS. Our QSDC protocol has longer maximal communication distance about 17.975 km with average photon number of 0.01. Our work serves as a
major step toward the further development of practical passive decoy-state QSDC systems.
\end{abstract}

\maketitle


\section{Introduction}
Quantum communication has emerged as a promising avenue, owing to its unconditional security guaranteed by the principles of quantum mechanics. The pioneering quantum communication protocol, known as quantum key distribution (QKD) \cite{bb84,Ekert91}, which can share random keys between two distant users. Over the past years, QKD has achieved significant advancements  in both theoretical \cite{qkd1,qkd2,qkd3,qkd4,qkd5,qkd6,qkd7} and experimental aspects \cite{exp1,exp2,exp3,exp4}. Another remarkable branch of quantum communication is the quantum secure direct communication (QSDC), which was first proposed by Long \cite{QSDC1}. Unlike QKD, QSDC allows the direct transmission of secure messages without sharing keys in advance. In 2003 and 2004, entanglement-based (two-step QSDC) and single-photon-based (DL04) QSDC protocols were proposed \cite{QSDC2,QSDC3}. The two-step QSDC protocol utilizes dense coding method, which can transmit two bits of messages with one pair of entangled Bell state. The DL04 QSDC protocol can transmit one bit of message using a single photon.  QSDC also has gained significant development in recent years \cite{QSDC5,QSDC9,QSDC10,QSDC11,DI,MDI2,QSDC14n,li,Masking,Mapping,QSDC12,QSDC13,QSDC15,QSDC16,QSDC16n,QSDC17n,QSDC17,QSDC18,QSDC19,QSDC20,QSDC20n,QSDC21,QSDC22,QSDC23,QSDC24}. The DL04 QSDC protocol and two-step QSDC protocol were realized in experiment in 2016 and 2017, respectively \cite{QSDC9,QSDC10}. Later, some pivotal QSDC experiments have been reported, such as the QSDC network experiment \cite{QSDC12}, 100-km fiber QSDC experiment \cite{QSDC15}, and continuous-variable QSDC experiment \cite{QSDC17,QSDC20n,Mapping}. In the theoretical aspect, the device-independent (DI) QSDC protocol and measurement-device-independent (MDI) QSDC protocol have been proposed to resist the potential attacks focusing on imperfect experimental equipment \cite{DI,MDI2,QSDC19}. Recently, one-step QSDC based on polarization-spatial-mode
hyperentanglement was proposed \cite{QSDC13}. Compared to traditional two-step QSDC, the one-step QSDC protocol requires only one round of photon transmission, which can simplify the experimental operation and significantly reduce the message loss caused by the photon transmission loss. Later, the DI one-step QSDC protocol \cite{QSDC16} and MDI one-step QSDC protocol \cite{QSDC16n} were successively proposed, which can enhance one-step QSDC's security under practical experimental condition.

Single-photon source plays a role in both QKD and QSDC \cite{bb84,QSDC3}. Unfortunately, ideal single-photon source is not available under current experimental conditions. The current available single-photon source is the phase-randomized weak coherent pulse
(WCP) source, which can emit vacuum state, single-photon state, and multiphoton state with different probabilities. Imperfect single-photon source not only decreases the communication efficiency of quantum communication, but also introduces a security loophole. An alternative approach is to use the device-independent (DI) type protocols, i.e., DI-QKD protocols \cite{qkd1,DIQKD,DIQKD1,DIQKD2} and DI-QSDC protocols \cite{DI,QSDC16,QSDC18}. However, these protocols are limited by the short communication distance. Moreover, they are also hard to realize in experiment. DI-QKD has been experimentally demonstrated until 2022 \cite{DIE1,DIE2,DIE3} and DI-QSDC has not obtained experimental demonstration yet.  Another alternative approach is to use the heralded single-photon source (HSPS) \cite{HSPS1,HSPS2,HSPS3,HSPS4,HSPS5,HSPS6,HSPS7}. Suppose that a spontaneous parametric down-conversion (SPDC) source emits correlated photon pairs in two spatial modes. By detecting the photons in one of the two correlated spatial modes, it is possible to  infer the photon-number statistics of the other spatial mode. This approach can significantly reduce the probability of the vacuum state occurrence. Furthermore, the  multiphoton components in the WCP may provide the eavesdropper (Eve) an opportunity to exploit the photon-number-splitting (PNS) attack \cite{PNS1,PNS2}. Fortunately, the PNS attack in QKD and QSDC can be effectively resisted by the decoy-state methods \cite{decoy0,decoy,decoy1,decoy3,decoy4}. However, the decoy-state method  necessitates active modulation of light intensity, which could potentially be exploited by Eve for a side-channel attack. Passive protocols are more
resistant to side-channel attacks than active systems \cite{passive1,passive2,passive3,passive4,passive5,passive6,passive7,passive8}. In the passive protocols, two phase-randomized WCPs interfere at a beam splitter (BS), which makes the photon-number statistics of the
outcome signals classically correlated. By measuring one of the two outcome signals of the BS,
we can passively obtain the conditional photon-number distribution of the other signal mode.

 In this paper, we introduce the passive decoy-state method and the HSPS into the single-photon-based QSDC, and propose a passive decoy-state QSDC protocol with the HSPS. Comparing with the original DL04 QSDC protocol with the WCP source, this passive decoy-state QSDC protocol has two attractive advantages. First, using the passive decoy-state method, this QSDC protocol has strong robustness against the side-channel attack. Second, with HSPS, this protocol can largely reduce the influence from the vacuum state, and thus the secrecy message capacity and maximal communication distance can be increased. This paper is organized as follows. In Sec. II, we explain the passive decoy-state QSDC protocol with the HSPS in detail. In Sec. III, we analyze its secrecy message capacity under practical experimental conditions in theory and perform the numerical simulation. In Sec. IV, we provide a conclusion.

\section{The passive decoy-state QSDC protocol with the HSPS}
\begin{figure*}[!htbp]
	\begin{center}
		\includegraphics[width=15cm,angle=0]{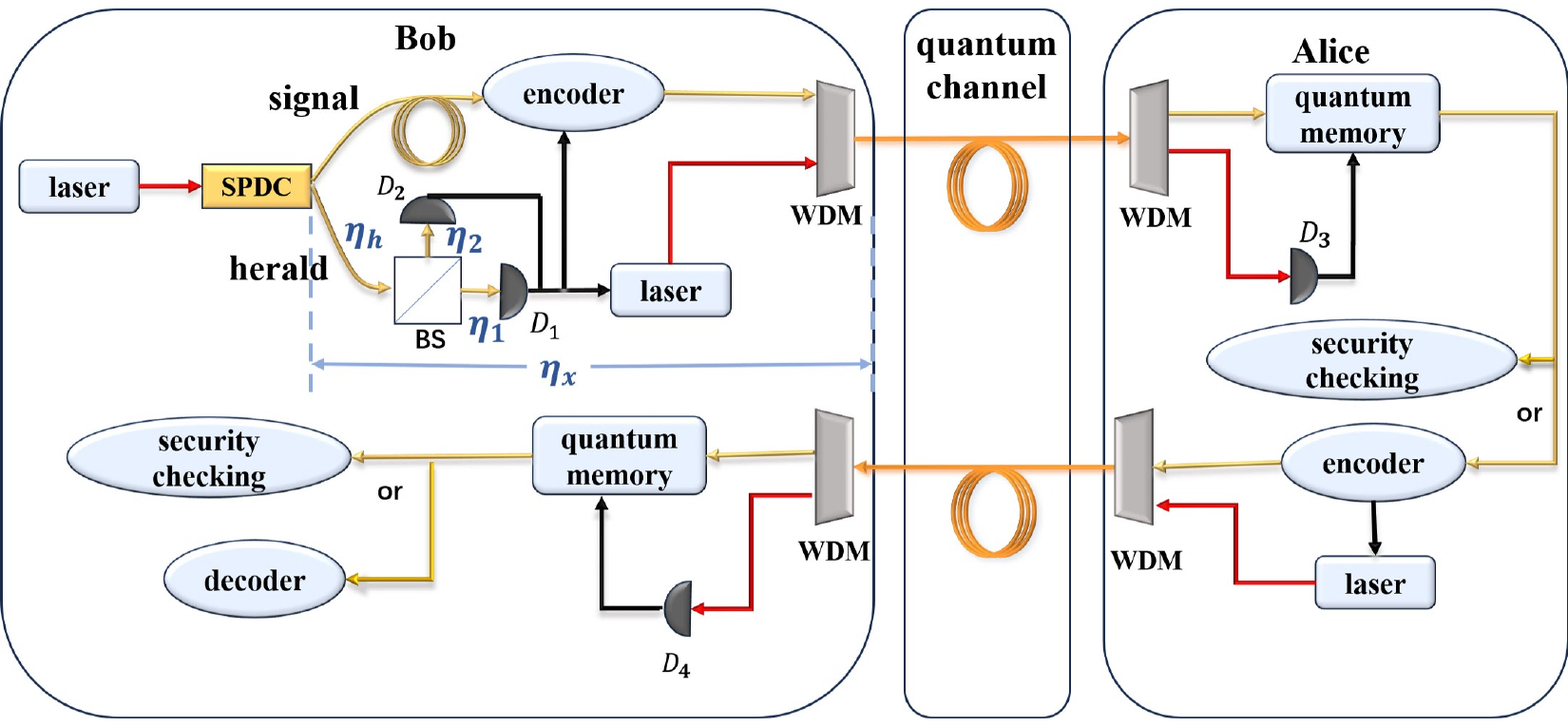}
		\caption{The structure of the passive decoy-state QSDC protocol with the HSPS. In the diagram, the yellow lines represent the spatial paths of the photons, while the orange lines represent the quantum channels used for communication. The red lines indicate the input laser, which is used as a source to pump a nonlinear crystal or a calibration tool to assist Alice's and Bob's memory devices to store photons. The black lines represent the detector response signals, which prompt Alice or Bob to perform a series of operations, including encoding, decoding, storing, and sending the indication laser. Overall, this process involves two rounds of security checking to ensure the security and integrity of the transmitted messages.}\label{simulation}
	\end{center}
\end{figure*}

In this section, we explain our passive decoy-state QSDC protocol with the HSPS.  As shown in  Fig. 1, the passive decoy-state QSDC protocol can be described as follows.

Step 1: The message receiver Bob passes a WCP in $|H\rangle$ to pump an SPDC crystal, splitting a single photon to two correlated photons in $|HH\rangle$ probabilistically. One  photon is in the heralded path and the other photon is  in the signal path. The photon  in the heralded path passes through a $t:(1-t)$ BS and the output photon will be detected by two photon detectors $D_{1}$ and $D_{2}$. In this way, the responses of $D_{1}$ and $D_{2}$ will herald the existence of the photon in the signal path. In detail, when only $D_{1}$ or $D_{2}$ responds, the pulse in the signal path will be used as signal states. When $D_{1}$ and $D_{2}$ both respond, the pulse in the signal path will be used as the decoy state. When neither $D_{1}$ nor $D_{2}$ clicks, the pulse in the signal path would be discarded.

Step 2:  Based on the detector responses, Bob performs the randomly encoding operation to generate one of the four polarization states, i. e., $|H\rangle$, $|V\rangle$, $|+\rangle$, $|-\rangle$. Here, $|H\rangle$ (horizontal polarization) and $|V\rangle$ (vertical polarization) belong to the rectilinear ($Z$) basis and $|+\rangle=\frac{1}{\sqrt{2} } (| H  \rangle+| V   \rangle)$ and $| -\rangle = \frac{1}{\sqrt{2} } (| H  \rangle-| V   \rangle)$ belong to the diagonal ($X$) basis.
  Bob sends the randomly encoded signal photons and decoy-state photons to the message sender Alice through the quantum channel. Simultaneously, Bob also sends an indication laser pulse  to prompt Alice to perform the storage operation.

Step 3: Alice stores the signal states and decoy states in the quantum memory if she receives the indication laser pulse. Then, she performs the first round of security checking, which is similar as that of the decoy-state QKD protocols \cite{decoy0,decoy}. After the security checking, the communication parties calculate the quantum bit error rate $E1$. If $E1$ is below the tolerate threshold, the communication continues. Otherwise, the parties have to discard the communication.

Step 4: Alice extracts the remaining signal photons from the quantum memory and encodes her messages by performing $I$ or $Y$ operation on each photon, where $I=|H\rangle\langle H|+|V\rangle\langle V|$ and $Y=i\sigma_{y}=|H\rangle\langle V|-|V\rangle\langle H|$. $I$ and $Y$ represent the classical messages 0 and 1, respectively. Meanwhile, Alice also randomly encodes a part of photons as the second round of security-checking photons. After encoding, Alice sends all the encoded photons to Bob through the quantum channel. Meanwhile, she sends another indication laser pulse to ask Bob to receive the signal states.

Step 5: After receiving the transmitted photons, Bob stores the photons in quantum memory and Alice announces the positions and encoding operations of the security-checking photons. Bob extracts the security-checking photons and measures them with the same basis as he prepared the initial photons. After the measurement, Bob calculates the quantum bit error rate $E2$ combined with Alice's announcement and his initial photon states. If $E2$ is below the threshold, the communication continues. Otherwise, the parties have to discard the communication.

Step 6: Bob extracts the remaining signal photons from the quantum memory and measures the photons with the same basis in which he prepared the initial photons. After the measurements, Bob can deduce Alice's operation by comparing the initial quantum state and the encoded quantum state, and thus obtain Alice's transmitted messages.

As shown in Fig. 1, in step 1, we suppose that the SPDC source emits two photon pulses in the signal path and heralded path with the probability of $P_{\mu }(k)$, which satisfies the Poisson distribution ($\mu$ is the average photon number) \cite{spdc1,spdc2,spdc3}. There are four possible combinations of detector responses after the photons in the heralded path passing through the $t:(1-t)$ BS. We note them as four events $x_{i}$ $(i=1,2,3,4)$, where $x_{1}=\bar{D} _{1}\bar{D} _{2}$, $x_{2}=D _{1}\bar{D} _{2}$, $x_{3}=\bar{D} _{1}D _{2}$, $x_{4}=D _{1}D _{2}$. $D_{j}$ represents the photon detector $D_j$ has a response and $\bar{D}_{j}$ represents that the detector $D_j$ has no response. We define $\gamma_{x_{i}}(k)$ as the probability of $k$ photons in the heralded path leading to the event $x_{i}$.
Hence, in the signal path, the photon-number distribution function $q_{x_{i}}(n)$ after the heralded generation is
\begin{eqnarray}\label{pnd}
	q_{x_{i}}(n)  &=&  \sum_{k  =  n}^{\infty } P_{\mu } (k)\gamma_{x_{i}} (k)S(k,n),   \nonumber\\
P_{\mu } (k) & = & e^{-\mu }\frac{\mu ^{k}}{k!}, \nonumber\\
	S(k,n)&=& C^{n}_{k}\eta_{x}^{n}(1-\eta_{x})^{k-n}.
\end{eqnarray}
Here, $S(k,n)$ is the probability that the emitted $k$ photons turn to $n$ photons due to the photon loss with the transmission efficiency of $\eta_{x}$. This process can be understood as the source emits $k$ photons which are successfully detected in the heralded path. Then, after a series of losses, the photon number in the signal path is reduced to $n$.

$\gamma_{x_{i}} (k)$ ($i=1,2,3,4$) varies depending on the responses of the detectors, which is shown as
\begin{eqnarray}
	\gamma _{x_{1}}(k)&=& (1-d_{1})(1-d_{2})f^{k},\nonumber \\
	\gamma _{x_{2}}(k)&=&(1-d_{2})f_{1}^{k}-\gamma _{x_{1}}(k), \nonumber \\
	\gamma _{x_{3}}(k)&=&(1-d_{1})f_{2}^{k}-\gamma _{x_{1}}(k), \nonumber \\
	\gamma _{x_{4}}(k)&=&1-\gamma _{x_{1}}(k)-\gamma _{x_{2}}(k)-\gamma _{x_{3}}(k).\label{gamma}
\end{eqnarray}
Here $f$, $f_{1}$ and $f_{2}$ represent the photon-loss probabilities in the heralded path, while $d_{1}$ ($d_{2}$) is the dark count rates of Alice's detector $D_{1}$ ($D_{2}$).

We define the coefficients $\eta _{h}$ as the transmission efficiency in the heralded path, and define $\eta _{1}$  ($\eta _{2}$)  as the detection  efficiency of the detector $D_{1}$ ($D_{2}$). In this way, we can obtain $f$, $f_{1}$ and $f_{2}$ as
\begin{eqnarray}
	f&=&\eta _{h}[t(1-\eta_{1})+(1-t)(1-\eta_{2})]  +1-\eta_{h}, \nonumber \\
	f_{1}&=&\eta _{h}[1-(1-t)\eta_{2}]+1-\eta_{h},\nonumber \\
	f_{2}&=&\eta _{h}(1-t\eta_{1})+1-\eta_{h} .\label{f}
\end{eqnarray}
By substituting the parameters in Eq. (\ref{gamma}) and Eq. (\ref{f}) into Eq. (\ref{pnd}), we can  further  derive $q_{x_{i}}(n)$ $(i=1,2,3,4)$ as
\begin{eqnarray}\label{pds}
	q_{x_{1}}(n)&=&(1-d_{1})(1-d_{2})\frac{(\mu \eta_{x}f)^{n}}{n!} e^{\mu [f(1-\eta_{x})-1]},  \nonumber \\
	q_{x_{2}}(n)&=&(1-d_{2})\frac{(\mu \eta_{x}f_{1})^{n}}{n!} e^{\mu [f_{1}(1-\eta_{x})-1]}-q_{x_{1}}(n),  \nonumber\\
	q_{x_{3}}(n)&=&(1-d_{1})\frac{(\mu \eta_{x}f_{2})^{n}}{n!} e^{\mu [f_{2}(1-\eta_{x})-1]}-q_{x_{1}}(n), \nonumber \\
	q_{x_{4}}(n)&=&\frac{(\mu \eta_{x})^{n}}{n!} e^{-\mu \eta_{x}}-q_{x_{1}}(n)-q_{x_{2}}(n)-q_{x_{3}}(n).
\end{eqnarray}

Here, the detector response $x_{1}$ indicates the heralded failure. In this case, Bob will discard the photon pulse in the signal path, so that $q_{x_{1}}(n)$ should also be discarded. The detector responses $x_{2}$ and $x_{3}$ enable Bob to obtain two different distributed signal states $q_{x_{2}}(n)$ and $q_{x_{3}}(n)$ in the signal path. If the detector response $x_{4}$ is obtained, the pulse in the signal path is used as the decoy state. It can be found that by adopting the heralded generation method, the ratio corresponding to the vacuum state event in the signal path can be greatly reduced, which can effectively reduce the influence of the dark count on the photon-number distribution in the signal path.

\section{Security analysis}
\subsection{The theoretical secrecy message capacity of the single-photon-based QSDC }
According to  Wyner's wiretap channel theory \cite{wyner}, the secrecy message capacity of the single-photon-based QSDC can be calculated as
\begin{eqnarray}\label{Cs}
	C_{s}= \max_{P_{A}}\{I(A:B)-I(A:E) \} ,
\end{eqnarray}
where $I(A:B)$ is the mutual information between Alice and Bob, and $I(A:E) $ is the mutual information between Alice and Eve. We define $P_{A}$ as
the probability distribution of Alice's encoding operations. In general, we consider the case that the messages 0 and 1  sent by Alice are equally distributed, that is, $P(A0)=P(A1)=0.5$.

 In the single-photon-based QSDC protocol, two rounds of quantum transmission are needed. We denote the photon transmission from  Bob to Alice as $BA$, and the photon transmission from Bob to Alice and then back to Bob after Alice's encoding as $BAB$. $I(A:B) $ can be calculated as
\begin{eqnarray}
	I(A:B)= Q_{\mu }^{BAB} [1-h(E_{\mu}^{BAB} )],
\end{eqnarray}
where $Q_{\mu }^{BAB}$ is  the overall gain of a photon traveling from Bob to Alice and then back to Bob, and $E_{\mu}^{BAB} $ is the total QBER after two rounds of photon transmission.  The function $h(x)$ is the binary Shannon entropy with the form of $h(x)=-x\log_{2}(x)-(1-x)\log_{2}(1-x)$.

The mutual information $I(A:E)$ between Alice and Eve can be considered as the message leakage rate through imperfect quantum channels and devices. Due to the inherent characteristics of QSDC, Eve can steal the encoded messages only when he can steal the corresponding photons in both photon-transmission processes. If Eve  steals only some encoded photons  in the second round of photon transmission, he is unable to decode the encoded messages without knowing the initial quantum states of the photons.
Consequently, Eve's message interception rate  is upper bounded by his photon interception rate in the first photon-transmission round.

For Eve, there are many strategies he can adopt to maximize his photon interception rate. We first consider the one-photon case, say, the signal pulse contains exactly one photon. Here, we consider a common approach, say, the collective attack, which has been considered in many works \cite{collective1,collective2,decoy3}.
In the collective attack, Eve sets up an auxiliary quantum system and makes a joint operation on his intercepted photon and the auxiliary system.
It is generally assumed that Eve can perform an optimal unitary operation to maximize the amount of messages he can steal. According to Holevo's theorem \cite{holevo}, we can deduce that the maximal photon interception rate that Eve can obtain from a single photon pulse is $h(e_{X}^{BA}+e_{Z}^{BA})$, where  $e_{X}^{BA}$ and $e_{Z}^{BA}$ represent the error rates in the $X$ basis and $Z$ basis after the first photon-transmission process, respectively. From the analysis in Ref. \cite{decoy3}, if the signal pulse contains two or more photons, Eve can perform the PNS attack and the collective attack. In detail, when the signal pulse contains two photons, Eve steals one photon with the PNS attack, and performs a collective attack on the other photon. The maximal message interception rate that Eve can obtain from the two photon pulse can be calculated as $\frac{1}{2}h(2e_{2}^{BA})+\frac{1}{2}$ with the Holevo bound. When the signal pulse contains three or more photons, the photons emitted by Bob can be unambiguously discriminated by Eve \cite{unam}, and thus the encoded messages can be completely stolen. As a result, $I(A:E) $ can be calculated as
\begin{eqnarray}\label{IAE}
	I(A:E)&=& \sum_{n=0}^{\infty } Q_{\mu,n }^{BAE}*H_{n}  \nonumber\\
	&=& Q_{\mu,n=1 }^{BAE}*h(2e_{1}^{BA} )+Q_{\mu,n=2 }^{BAE}*[\frac{1}{2} h(2e_{2}^{BA} )  \nonumber\\
	&&+\frac{1}{2}]+Q_{\mu,n\ge 3 }^{BAE}*1,
\end{eqnarray}
where $Q_{\mu,n}^{BAE}$ is the gain of the $n$-photon event from Eve, and $H_{n}$ is the contribution of the $n$-photon event to $I(A:E)$. The one-photon event's contribution for $H_{n}$ is $h(e_{X}^{BA}+e_{Z}^{BA})$. Here, we assume that $e_{X}^{BA}$=$e_{Z}^{BA}$=$e_{1}^{BA}$. $e_{2}^{BA}$ represents the total error rate after the first photon transmission caused by the two-photon event.

In this way, we can obtain the secrecy message capacity as
\begin{eqnarray}\label{C}
	C_{s}&=& Q_{\mu }^{BAB} [1-h(E_{\mu}^{BAB} )]- \{Q_{\mu,n=1 }^{BAE}*h(2e_{1}^{BA} )+\nonumber\\
	&&Q_{\mu,n=2 }^{BAE}*[\frac{1}{2} h(2e_{2}^{BA} )  +\frac{1}{2}]+Q_{\mu,n\ge 3 }^{BAE}*1\}.
\end{eqnarray}
\subsection{System model}
In order to analyze the secrecy message capacity of our passive decoy-state QSDC protocol with the HSPS, we establish a QSDC system simulation model, including the source, channel, detector, and yield.

 As shown in Sec. II,  Eq. (\ref{pds}) provides the photon-number distributions of two signal sources ($q_{x_{2}}$ and $q_{x_{3}}$) and a decoy-state source ($q_{x_{4}}$). The channel transmission efficiency is shown as
\begin{eqnarray}
	t^{chan}&=&10^{-\frac{\alpha ^{chan}}{10} }, 
\end{eqnarray}
where $\alpha ^{chan}$ is the loss of quantum channel and chan$\in\{BA,BAB\}$.

In this situation, the overall transmission efficiency $\eta ^{chan}$ of the signal state and decoy state can be expressed as
\begin{eqnarray}
	\eta ^{chan}&= &t^{chan}\eta_{opt} ^{chan}\eta_{d} ^{par}, 
\end{eqnarray}
where $ \eta_{opt} ^{chan}$ is the intrinsic optical efficiency of the device, and $\eta_{d} ^{par}$ is the detection efficiency of Alice or Bob (par$\in \{A,B\}$).

In order to calculate the secrecy message capacity, we need to obtain the yield and gain of the channel.
Let $Y_{n}^{A}$ and $Y_{n}^{B}$ denote the yields of the $n$-photon signal at Alice's and Bob's locations, respectively. They can be calculated as
\begin{eqnarray}
	Y_{n}^{par}=1-(1-Y_{0}^{par})(1-\eta ^{chan})^{n},
\end{eqnarray}
where $Y_{0}^{par}$ is the background detection rate. The item $(1-Y_{0}^{par})(1-\eta ^{chan})^{n}$ can be understood as the probability that no background detection event occurs and all the $n$ photons are lost in the quantum channel.

Through the system model described above, we can estimate both the overall gain and the error rate. The formula of the overall gain can be written as
\begin{eqnarray}
 Q_{x_{i} }^{chan}=\sum_{n=0}^{\infty } Q_{x_{i},n }^{chan}=\frac{1}{P_{x_{i}}}\sum_{n=0}^{\infty }q_{x_{i}}(n)Y_{n}^{par},
\end{eqnarray}
  where $P_{x_{i}} $ is the total probability of event $x_{i}$ and $P_{x_{i}} =\sum_{n=0}^{\infty } q_{x_{i}}(n) $.
  We can further derive $Q_{x_{1} }^{chan}$ as
  \begin{eqnarray}
	&&Q_{x_{1} }^{chan}=\sum_{n=0}^{\infty } Q_{x_{1},n }^{chan}=\frac{1}{P_{x_{1}}}\sum_{n=0}^{\infty }q_{x_{1}}(n)Y_{n}^{par}\\ \nonumber
	&=&\frac{1}{P_{x_{1}}}\sum_{n=0}^{\infty }q_{x_{1}}(n)[1-(1-Y_{0}^{par})(1-\eta^{chan})^n]\\ \nonumber
	&=&\frac{1}{P_{x_{1}}} \sum_{n=0}^{\infty }q_{x_{1}}(n)-\frac{1}{P_{x_{1}}}\sum_{n=0}^{\infty }q_{x_{1}}(n)(1-Y_{0}^{par})(1-\eta^{chan})^n\\ \nonumber
	&=&1-\frac{1}{P_{x_{1}}}\sum_{n=0}^{\infty }(1-d_{1})(1-d_{2})\frac{(\mu \eta_{x}f)^{n}}{n!} e^{\mu [f(1-\eta_{x})-1]}\\\nonumber
	&\ast&(1-Y_{0}^{par})(1-\eta^{chan})^n \\\nonumber
	&=&1-\frac{1}{P_{x_{1}}}\sum_{n=0}^{\infty }(1-d_{1})(1-d_{2})\frac{[\mu \eta_{x}f(1-\eta^{chan})]^{n}}{n!}\\ \nonumber
	&*&e^{-\mu \eta_{x}f(1-\eta^{chan})} e^{\mu [f(1-\eta_{x})-1]+\mu \eta_{x}f(1-\eta^{chan})}(1-Y_{0}^{par}) \\\nonumber
	&=&1-\frac{1}{P_{x_{1}}}(1-Y_{0}^{par})(1-d_{1})(1-d_{2}) e^{-\mu\eta_{x}f\eta^{chan} +\mu f-\mu}.
\end{eqnarray}
For simplicity, we rewrite
\begin{eqnarray}
Q_{x_{1} }^{chan}&=&1-\frac{1}{P_{x_{1}}}(1-Y_{0}^{par})(1-d_{1})(1-d_{2}) g(f), \\
	g(f)&=&e^{-\mu f\eta_{x} \eta ^{chan}+\mu f-\mu },
\end{eqnarray}
where $g(f)$ is a substitute function for aesthetics. In addition, note that $Q_{x_{1} }^{chan}$ here is only a computational example, and we do not need to calculate $Q_{x_{1} }^{chan}$ in practical experiment or simulations. Similarly, we can obtain the overall signal gains of the three kinds of heralded states (two signal states and one decoy state) as
\begin{eqnarray}\label{Q}
	 Q_{x_{2} }^{chan}&=&1-\frac{1}{P_{x_{2}}}(1-Y_{0}^{par})(1-d_{2})[g(f_{1})-(1-d_{1})g(f)],\nonumber\\
	Q_{x_{3} }^{chan}&=&1-\frac{1}{P_{x_{3}}}(1-Y_{0}^{par})(1-d_{1})[g(f_{2})-(1-d_{2})g(f)],\nonumber\\
	Q_{x_{4} }^{chan}&=&1-\frac{1}{P_{x_{4}}}(1-Y_{0}^{par})[g(1)-(1-d_{1})g(f_{2})\nonumber\\
	&&-(1-d_{2})g(f_{1})+(1-d_{1})(1-d_{2})g(f)].
\end{eqnarray}

 According to Ref. \cite{decoy3}, the overall signal gain of Eve can be calculated as
\begin{eqnarray}
	Q_{x_{i}}^{BAE}&=&\sum_{n=0}^{\infty } Q_{x_{i},n }^{BAE}\\\nonumber
	&\le& \sum_{n=0}^{\infty } [Q^{BA}_{x_{i} ,n}-\frac{q_{x_{i} }(n)}{P_{x_{i}}}Y_{0}^{A}]max\{1,\frac{\gamma ^{E}}{\gamma^{A}}\},
\end{eqnarray}
where $\gamma ^{E}$  is the overall transmission efficiency of Eve after Alice encodes her receiving photons, and $\gamma^{A}$ is the overall transmission efficiency for photons received and then measured by Alice.

Similarly, we can also calculate the total error rate of our QSDC protocol as
\begin{eqnarray}
E_{x_{i} }^{chan}=\frac{\sum_{n=0}^{\infty }q_{x_{i}}(n)e_{n}Y_{n}^{par}}{Q_{x_{i} }^{chan}P_{x_{i}}}.
\end{eqnarray}
Here, we construct the error model $e_{n}Y_{n}^{par} $ as
\begin{eqnarray}
	e_{n}Y_{n}^{par}=e_{0}^{par}Y_{0}^{par}+e_{d}^{par}[1-(1-\eta)^n],
\end{eqnarray}
where $e_{d}^{par}$ is detector error rate and $e_{0}^{par}$ is the error rate caused by the dark count. $e_{0}^{par}$ is equal to 0.5, which means when no photon  arrives, the dark count from one of the two detectors may cause the error with the probability of 0.5.

Here, we also take the calculation of the total error rate $E_{x_{1} }^{chan}$ as an example. In detail, we can obtain
  \begin{eqnarray}
	&&E_{x_{1} }^{chan}=\frac{\sum_{n=0}^{\infty }q_{x_{1}}(n)e_{n}Y_{n}^{par}}{Q_{x_{1} }^{chan}P_{x_{1}}} \\
	&=&\frac{1}{Q_{x_{1} }^{chan}P_{x_{1}}}\sum_{n=0}^{\infty }q_{x_{1}}(n)\{e_{0}^{par}Y_{0}^{par}+e_{d}^{par}[1-(1-\eta)^n]\} \nonumber\\
	&=&\frac{1}{Q_{x_{1} }^{chan}P_{x_{1}}}[\sum_{n=0}^{\infty }q_{x_{1}}(n)(e_{0}^{par}Y_{0}^{par}+e_{d}^{par})\nonumber \\
	&-&\sum_{n=0}^{\infty }q_{x_{1}}(n)e_{d}^{par}(1-\eta)^n]  \nonumber\\ \nonumber
	&=&\frac{e_{0}^{par}Y_{0}^{par}+e_{d}^{par}}{Q_{x_{1} }^{chan}}-\frac{e_{d}^{par}}{Q_{x_{1} }^{chan}P_{x_{1}}}(1-d_{1})(1-d_{2}) g(f).
\end{eqnarray}

Similarly, we can calculate the total error rate of the two signal states and one decoy state as
\begin{eqnarray}\label{E}
	E_{x_{2} }^{chan}&=&\frac{1}{Q_{x_{2} }^{chan}}(e_{0}^{par}Y_{0}^{par}+e_{d}^{par}) \nonumber\\
&&-\frac{e_{d}^{par}}{Q_{x_{2} }^{chan}P_{x_{2}}}(1-d_{2})[g(f_{1})-(1-d_{1})g(f)], \nonumber\\
	E_{x_{3} }^{chan}&=&\frac{1}{Q_{x_{3} }^{chan}}(e_{0}^{par}Y_{0}^{par}+e_{d}^{par})\nonumber\\
&&-\frac{e_{d}^{par}}{Q_{x_{3} }^{chan}P_{x_{3}}}(1-d_{1})[g(f_{2})-(1-d_{2})g(f)], \nonumber\\
	E_{x_{4} }^{chan}&=&\frac{1}{Q_{x_{4} }^{chan}}(e_{0}^{par}Y_{0}^{par}+e_{d}^{par})\nonumber\\
&&-\frac{e_{d}^{par}}{Q_{x_{4} }^{chan}P_{x_{4}}}[g(1)-(1-d_{1})g(f_{2})\nonumber\\
&&-(1-d_{2})g(f_{1})+(1-d_{1})(1-d_{2})g(f)].
\end{eqnarray}

In this way, we can estimate $I(A:B)$ based on Eq. (\ref{Q}) and Eq. (\ref{E}).

According to Eq. (\ref{IAE}), for estimating $I(A:E)$, we need to estimate the single-photon error rate $e_{1}^{BA}$ and two-photon bit error rate $e_{2}^{BA}$. Here, we adopt the decoy-state method to estimate $e_{1}^{BA}$ and $e_{2}^{BA}$.
In Sec. II, we generate three types of heralded states through the heralded operation. After Alice receives the transmitted photons, Bob announces the type of each photon pulse, and selects a part of pulses for security checking. If Eve performs the PNS attack during the photon transmission, he will inevitably change the photon distribution, and thus be detected by the parties.
After the first round of security checking, we have
\begin{eqnarray}\label{PQ}
  P_{x_{2}}Q_{x_{2} }^{BA}&=&q_{2}^{0}Y_{0}^{A}+q_{2}^{1}Y_{1}^{A}+q_{2}^{2}Y_{2}^{A}+\sum_{n=3}^{\infty }q_{2}^{n}Y_{n}^{A}, \nonumber\\
	P_{x_{3}}Q_{x_{3} }^{BA}&=&q_{3}^{0}Y_{0}^{A}+q_{3}^{1}Y_{1}^{A}+q_{3}^{2}Y_{2}^{A}+\sum_{n=3}^{\infty }q_{3}^{n}Y_{n}^{A},\nonumber\\
	P_{x_{4}}Q_{x_{4} }^{BA}&=&q_{4}^{0}Y_{0}^{A}+q_{4}^{1}Y_{1}^{A}+q_{4}^{2}Y_{2}^{A}+\sum_{n=3}^{\infty }q_{4}^{n}Y_{n}^{A},
\end{eqnarray}
\begin{eqnarray}\label{PQE}
	P_{x_{2}}Q_{x_{2} }^{BA}E_{x_{2} }^{BA}&=&q_{2}^{0}e_{0}Y_{0}^{A}+q_{2}^{1}e_{1}Y_{1}^{A}+q_{2}^{2}e_{2}Y_{2}^{A}\nonumber\\
&	+&q_{2}^{3}e_{3}Y_{3}^{A}+\sum_{n=4}^{\infty }q_{2}^{n}e_{n}Y_{n}^{A},\nonumber\\
	P_{x_{3}}Q_{x_{3} }^{BA}E_{x_{3} }^{BA}&=&q_{3}^{0}e_{0}Y_{0}^{A}+q_{3}^{1}e_{1}Y_{1}^{A}+q_{3}^{2}e_{2}Y_{2}^{A}\nonumber\\
&	+&q_{3}^{3}e_{3}Y_{3}^{A}+\sum_{n=4}^{\infty }q_{3}^{n}e_{n}Y_{n}^{A},\nonumber\\
	P_{x_{4}}Q_{x_{4} }^{BA}E_{x_{4} }^{BA}&=&q_{4}^{0}e_{0}Y_{0}^{A}+q_{4}^{1}e_{1}Y_{1}^{A}+q_{4}^{2}e_{2}Y_{2}^{A}\nonumber\\
&	+&q_{4}^{3}e_{3}Y_{3}^{A}+\sum_{n=4}^{\infty }q_{4}^{n}e_{n}Y_{n}^{A},
\end{eqnarray}
where $q_{i}^{n}$ is the short for $q_{x_{i}}(n)$. Note that the parameters used in the parameter estimation are all from the first round of photon transmission, and we have omitted the superscript $BA$ of $Q$ and $E$ below.

Here, $Y_{0}^{A}$ is the background detection rate which we consider as an inherent property of the detector. Then, based on Eq. (\ref{PQ}), we can  calculate $Y_{1}^{A}$ as
\begin{eqnarray}
	Y_{1}^{A}	&= &\frac{q_{i}^{2}P_{x_{j}}Q_{x_{j}}-q_{j}^{2}P_{x_{i}}Q_{x_{i}}-(q_{i}^{2}q_{j}^{0}-q_{j}^{2}q_{i}^{0})Y_{0}^{A}
	}{q_{i}^{2}q_{j}^{1}-q_{j}^{2}q_{i}^{1}}\nonumber\\
&-&\frac{\sum_{n=3}^{\infty }(q_{i}^{2}q_{j}^{n}-q_{j}^{2}q_{i}^{n})Y_{n}^{A}}{q_{i}^{2}q_{j}^{1}-q_{j}^{2}q_{i}^{1}}.
\end{eqnarray}
According to Ref. \cite{HSPS7}, we have
\begin{eqnarray}\label{e25}
	\frac{  q_{4}^{n}}{  q_{3}^{n}} \ge \frac{  q_{4}^{2}}{  q_{3}^{2}} \ge \frac{  q_{4}^{1}}{  q_{3}^{1}},\qquad
	\frac{  q_{3}^{n}}{  q_{2}^{n}} \ge \frac{  q_{3}^{2}}{  q_{2}^{2}} \ge \frac{  q_{3}^{1}}{  q_{2}^{1}}.
\end{eqnarray}
Eq. (\ref{e25}) can be further rewritten as
\begin{eqnarray}\label{e26}
	\frac{  q_{i}^{n}}{  q_{j}^{n}} \ge \frac{  q_{i}^{2}}{  q_{j}^{2}} \ge \frac{  q_{i}^{1}}{  q_{j}^{1}},\qquad
	i\ge j, \quad  i,j\in \{2,3,4\}.
\end{eqnarray}
According to Eq. (\ref{e26}), we have
\begin{eqnarray}
	\frac{(q_{i}^{2}q_{j}^{n}-q_{j}^{2}q_{i}^{n})}{q_{i}^{2}q_{j}^{1}-q_{j}^{2}q_{i}^{1}}\le 0.
\end{eqnarray}

As a result, we have the lower bound ($Y_{1}^{l}$) of $Y_{1}^{A}$ as
\begin{eqnarray}
	Y_{1}^{A}	&\ge &\frac{q_{i}^{2}P_{x_{j}}Q_{x_{j}}-q_{j}^{2}P_{x_{i}}Q_{x_{i}}-(q_{i}^{2}q_{j}^{0}-q_{j}^{2}q_{i}^{0})Y_{0}^{A}}{q_{i}^{2}q_{j}^{1}-q_{j}^{2}q_{i}^{1}},\nonumber\\
	Y_{1}^{l}&=&   \! \! \!  \max_{i\ge j \&  i,  j \atop \in \{2,3,4\}}   \! \! \!
	 \{ \frac{q_{i}^{2}P_{x_{j}}Q_{x_{j}}-q_{j}^{2}P_{x_{i}}Q_{x_{i}}-(q_{i}^{2}q_{j}^{0}-q_{j}^{2}q_{i}^{0})Y_{0}^{A}}{q_{i}^{2}q_{j}^{1}-q_{j}^{2}q_{i}^{1}}\}.\nonumber\\
	\end{eqnarray}
where the superscript l means the lower bound.

Similarly, $Y_{2}^{A}$ can be calculated as
\begin{eqnarray}
	Y_{2}^{A}&= &\frac{q_{i}^{1}P_{x_{j}}Q_{x_{j}}-q_{j}^{1}P_{x_{i}}Q_{x_{i}}-(q_{i}^{1}q_{j}^{0}-q_{j}^{1}q_{i}^{0})Y_{0}^{A}}{q_{i}^{1}q_{j}^{2}-q_{j}^{1}q_{i}^{2}} \nonumber\\
	&-&\frac{\sum_{n=3}^{\infty }(q_{i}^{1}q_{j}^{n}-q_{j}^{1}q_{i}^{n})Y_{n}^{A}}{q_{i}^{1}q_{j}^{2}-q_{j}^{1}q_{i}^{2}}.
\end{eqnarray}
According to Eq. (\ref{e26}), we have
\begin{eqnarray}
	\frac{(q_{i}^{1}q_{j}^{n}-q_{j}^{1}q_{i}^{n})}{q_{i}^{1}q_{j}^{2}-q_{j}^{1}q_{i}^{2}}\le 0.
\end{eqnarray}

In this way, the lower bound ($Y_{2}^{l}$) of $Y_{2}^{A}$  is shown as
\begin{eqnarray}
	Y_{2}^{A}&\ge &\frac{q_{i}^{1}P_{x_{j}}Q_{x_{j}}-q_{j}^{1}P_{x_{i}}Q_{x_{i}}-(q_{i}^{1}q_{j}^{0}-q_{j}^{1}q_{i}^{0})Y_{0}^{A}}{q_{i}^{1}q_{j}^{2}-q_{j}^{1}q_{i}^{2}} ,\nonumber\\
	Y_{2}^{l}&=& \! \! \! \max_{i\ge j \&  i,  j \atop \in \{2,3,4\}}  \! \! \! \{\frac{q_{i}^{1}P_{x_{j}}Q_{x_{j}}-q_{j}^{1}P_{x_{i}}Q_{x_{i}}-(q_{i}^{1}q_{j}^{0}-q_{j}^{1}q_{i}^{0})Y_{0}^{A}}{q_{i}^{1}q_{j}^{2}-q_{j}^{1}q_{i}^{2}}\}. \nonumber\\
\end{eqnarray}

According to Eq. (\ref{PQE}), we can estimate $e_{1}^{BA}$ and $e_{2}^{BA}$ as
\begin{widetext}
\begin{eqnarray}
	 e_{1}^{BA}&=&\frac{(q_{4}^{3}q_{3}^{2}-q_{3}^{3}q_{4}^{2})P_{x_{2}}Q_{x_{2}}E_{x_{2}}+(q_{2}^{3}q_{4}^{2}-q_{4}^{3}q_{2}^{2})P_{x_{3}}Q_{x_{3}}E_{x_{3}}+(q_{3}^{3}q_{2}^{2}-q_{2}^{3}q_{3}^{2})P_{x_{4}}Q_{x_{4}}E_{x_{4}}-\sum_{n  = 4}^{\infty}f_{q1}(n)e_{n}Y_{n}^{A}}
	 {\{q_{4}^{3}(q_{3}^{2}q_{2}^{1}-q_{2}^{2}q_{3}^{1})+q_{2}^{3}(q_{4}^{2}q_{3}^{1}-q_{3}^{2}q_{4}^{1})+q_{3}^{3}(q_{2}^{2}q_{4}^{1}-q_{4}^{2}q_{2}^{1})\}Y_{1}^{A}}\nonumber\\
	 &&-\frac{\{q_{4}^{3}(q_{3}^{2}q_{2}^{0}-q_{2}^{2}q_{3}^{0})+q_{2}^{3}(q_{4}^{2}q_{3}^{0}-q_{3}^{2}q_{4}^{0})+q_{3}^{3}(q_{2}^{2}q_{4}^{0}-q_{4}^{2}q_{2}^{0})\}e_{0}Y_{0}^{A}}
	 {\{q_{4}^{3}(q_{3}^{2}q_{2}^{1}-q_{2}^{2}q_{3}^{1})+q_{2}^{3}(q_{4}^{2}q_{3}^{1}-q_{3}^{2}q_{4}^{1})+q_{3}^{3}(q_{2}^{2}q_{4}^{1}-q_{4}^{2}q_{2}^{1})\}Y_{1}^{A}}\nonumber\\
	&\le &\frac{(q_{4}^{3}q_{3}^{2}-q_{3}^{3}q_{4}^{2})P_{x_{2}}Q_{x_{2}}E_{x_{2}}+(q_{2}^{3}q_{4}^{2}-q_{4}^{3}q_{2}^{2})P_{x_{3}}Q_{x_{3}}E_{x_{3}}+(q_{3}^{3}q_{2}^{2}-q_{2}^{3}q_{3}^{2})P_{x_{4}}Q_{x_{4}}E_{x_{4}}}
	 {\{q_{4}^{3}(q_{3}^{2}q_{2}^{1}-q_{2}^{2}q_{3}^{1})+q_{2}^{3}(q_{4}^{2}q_{3}^{1}-q_{3}^{2}q_{4}^{1})+q_{3}^{3}(q_{2}^{2}q_{4}^{1}-q_{4}^{2}q_{2}^{1})\}Y_{1}^{l}}\nonumber\\
	 &&-\frac{\{q_{4}^{3}(q_{3}^{2}q_{2}^{0}-q_{2}^{2}q_{3}^{0})+q_{2}^{3}(q_{4}^{2}q_{3}^{0}-q_{3}^{2}q_{4}^{0})+q_{3}^{3}(q_{2}^{2}q_{4}^{0}-q_{4}^{2}q_{2}^{0})\}e_{0}Y_{0}^{A}}
	 {\{q_{4}^{3}(q_{3}^{2}q_{2}^{1}-q_{2}^{2}q_{3}^{1})+q_{2}^{3}(q_{4}^{2}q_{3}^{1}-q_{3}^{2}q_{4}^{1})+q_{3}^{3}(q_{2}^{2}q_{4}^{1}-q_{4}^{2}q_{2}^{1})\}Y_{1}^{l}}\nonumber\\
	&=&e_{1}^{u},
\end{eqnarray}
\begin{eqnarray} e_{2}^{BA}&=&\frac{(q_{4}^{3}q_{3}^{1}-q_{3}^{3}q_{4}^{1})P_{x_{2}}Q_{x_{2}}E_{x_{2}}+(q_{2}^{3}q_{4}^{1}-q_{4}^{3}q_{2}^{1})P_{x_{3}}Q_{x_{3}}E_{x_{3}}+(q_{3}^{3}q_{2}^{1}-q_{2}^{3}q_{3}^{1})P_{x_{4}}Q_{x_{4}}E_{x_{4}}-\sum_{n  =  4}^{\infty }f_{q2}(n)e_{n}Y_{n}^{A}}
	 {\{q_{4}^{3}(q_{3}^{1}q_{2}^{2}-q_{2}^{1}q_{3}^{2})+q_{2}^{3}(q_{4}^{1}q_{3}^{2}-q_{3}^{1}q_{4}^{2})+q_{3}^{3}(q_{2}^{1}q_{4}^{2}-q_{4}^{1}q_{2}^{2})\}Y_{2}^{A}} \nonumber\\
	 &&-\frac{\{q_{4}^{3}(q_{3}^{1}q_{2}^{0}-q_{2}^{1}q_{3}^{0})+q_{2}^{3}(q_{4}^{1}q_{3}^{0}-q_{3}^{1}q_{4}^{0})+q_{3}^{3}(q_{2}^{1}q_{4}^{0}-q_{4}^{1}q_{2}^{0})\}e_{0}Y_{0}^{A}}
	{\{	 q_{4}^{3}(q_{3}^{1}q_{2}^{2}-q_{2}^{1}q_{3}^{2})+q_{2}^{3}(q_{4}^{1}q_{3}^{2}-q_{3}^{1}q_{4}^{2})+q_{3}^{3}(q_{2}^{1}q_{4}^{2}-q_{4}^{1}q_{2}^{2})\}Y_{2}^{A}} \nonumber\\
	&\le &\frac{(q_{4}^{3}q_{3}^{1}-q_{3}^{3}q_{4}^{1})P_{x_{2}}Q_{x_{2}}E_{x_{2}}+(q_{2}^{3}q_{4}^{1}-q_{4}^{3}q_{2}^{1})P_{x_{3}}Q_{x_{3}}E_{x_{3}}+(q_{3}^{3}q_{2}^{1}-q_{2}^{3}q_{3}^{1})P_{x_{4}}Q_{x_{4}}E_{x_{4}}}
	{\{	 q_{4}^{3}(q_{3}^{1}q_{2}^{2}-q_{2}^{1}q_{3}^{2})+q_{2}^{3}(q_{4}^{1}q_{3}^{2}-q_{3}^{1}q_{4}^{2})+q_{3}^{3}(q_{2}^{1}q_{4}^{2}-q_{4}^{1}q_{2}^{2})\}Y_{2}^{l}} \nonumber\\
	 &&-\frac{\{q_{4}^{3}(q_{3}^{1}q_{2}^{0}-q_{2}^{1}q_{3}^{0})+q_{2}^{3}(q_{4}^{1}q_{3}^{0}-q_{3}^{1}q_{4}^{0})+q_{3}^{3}(q_{2}^{1}q_{4}^{0}-q_{4}^{1}q_{2}^{0})\}e_{0}Y_{0}^{A}}
	 {\{q_{4}^{3}(q_{3}^{1}q_{2}^{2}-q_{2}^{1}q_{3}^{2})+q_{2}^{3}(q_{4}^{1}q_{3}^{2}-q_{3}^{1}q_{4}^{2})+q_{3}^{3}(q_{2}^{1}q_{4}^{2}-q_{4}^{1}q_{2}^{2})\}Y_{2}^{l}} \nonumber\\
	&=&e_{2}^{u}.
\end{eqnarray}
\end{widetext}
Here, $f_{q1}(n)$ ($f_{q2}(n)$) is a coefficient of order $n$, and its value is proved to be positive \cite{HSPS7}. By eliminating the positive term through scaling, we can estimate the upper bound of $e_{1}^{BA} $ and $e_{2}^{BA} $
as $e_{1}^{u} $ and $e_{2}^{u} $, respectively.

\subsection{Numerical simulation}
\begin{figure}[!htbp]\label{dis}
	\begin{center}
		\includegraphics[width=9cm,angle=0]{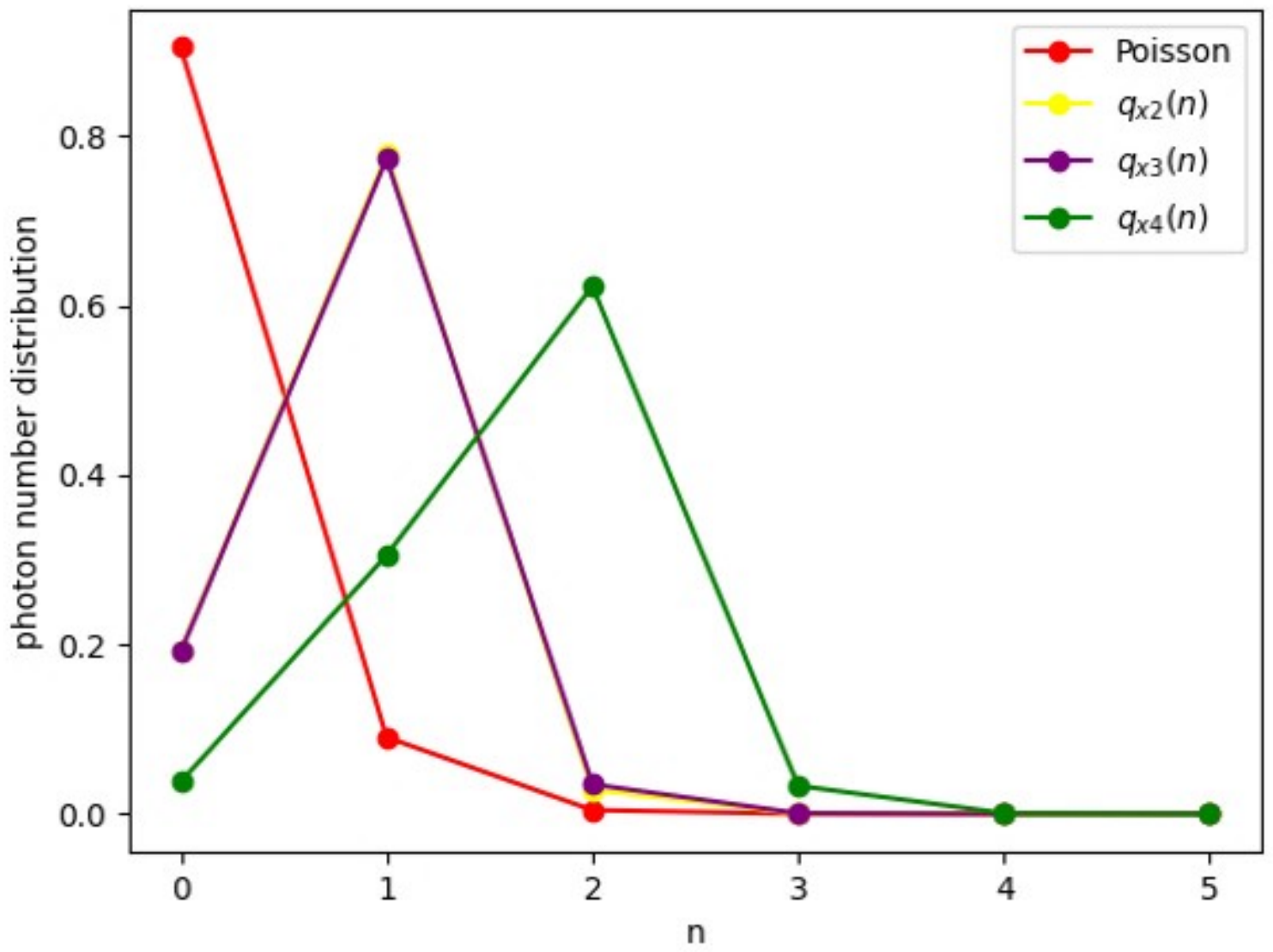}
		\caption{Comparison between the photon distribution of the HSPS and the Poisson distribution, where $\mu=0.1$  and the other parameters are shown in  Tab. 1. Notice that the curves $q_{x_{2}}(n)$ and $q_{x_{3}}(n)$ almost overlap. }\label{simulation}
	\end{center}
\end{figure}

\begin{table}[h]
    \centering
    \vspace{-0.1cm}
    \setlength{\abovecaptionskip}{0.3cm}
    \setlength{\belowcaptionskip}{0.1cm}
    \setlength\tabcolsep{5pt} 
    \renewcommand\arraystretch{1.5}  
    \caption{Parameters used in the numerical simulation. $\eta_{1},\eta_{2},\eta_{h},\eta_{x},t $ are the parameters that we model for the HSPS distribution, and the other parameters are from Pan's experiments \cite{decoy3}.}
    \begin{tabular}{cccccc}
       \hline
				\hline
        $\eta_{x}$ & $\eta_{1}$ & $\eta_{2}$ & $\eta_{h}$     & $\eta_{d}^{A(B)}$      &  t  \\
		                                          0.8                        & 0.6                        & 0.8                        & 0.9           &0.7 &0.4                  \\ \hline
		   $d_{1(2)} $                       &$Y_{0}^{A(B)} $  & $\eta_{opt}^{BA}$     & $\eta_{opt}^{BAB}$   & $e_{d}^{A}$                    & $e_{d}^{B}$                        \\
	       $8\times10^{-8}$                &      $8\times10^{-8}$                                                           &    0.21                        &        0.088                    &  0.0131                          &        0.0026                         \\ \hline 		\hline
    \end{tabular}
\end{table}

In our QSDC protocol, we utilize the HSPS to generate three types of sources with the photon-number distributions of $q_{x_{2}}(n)$, $q_{x_{3}}(n)$, and $q_{x_{4}}(n)$, corresponding to three response
events $x_{2}$, $x_{3}$, and $x_{4}$, respectively. Each of these three photon-number distributions as a function of the photon number $n$ is depicted in Fig. 2. The parameters used in the numerical simulation are shown in Tab. 1. Meanwhile, we also show the photon-number distribution of the WCP which follows the Poisson distribution. It can be found that the vacuum state accounts for a large proportion (90.4\%) in the WCP, while the
single-photon state  accounts only for 9.0\%. In contrast, by adopting the HSPS, the proportions of vacuum state in $q_{x_{2}}(n)$
and $q_{x_{3}}(n)$ largely reduce to about 19.3\% and 19.1\%, and the proportions of the single-photon state increase to about 77.8\% and 77.3\%, respectively, which will greatly benefit the gain.
Although the proportions of multiphoton events slightly increase in $q_{x_{2}}(n)$ and $q_{x_{3}}(n)$, their contribution can be negligible compared to the amplified single-photon component.
 The photon-number distribution $q_{x_{4}}(n)$ does not meet the ideal standard since the two-photon event accounts for a high proportion (about 60\%). However, we  use only $q_{x_{2}}(n)$ and
 $q_{x_{3}}(n)$ to transmit messages, but use $q_{x_{4}}(n)$ for the decoy state.

Therefore, the secrecy message capacity of our passive decoy-state QSDC protocol with the HSPS is written as
\begin{eqnarray}\label{C}
	C_{s}&=&C_{sq_{x_{2}}}+C_{sq_{x_{3}}}\\
	&=& Q_{x_{2} }^{BAB} [1-h(E_{x_{2}}^{BAB} )]- \{Q_{x_{2},n=1 }^{BAE}*h(2e_{1}^{BA} )\nonumber\\
	&&+Q_{x_{2},n=2 }^{BAE}*[\frac{1}{2} h(2e_{2}^{BA} )  +\frac{1}{2}]+Q_{x_{2},n\ge 3 }^{BAE}*1\}\nonumber\\
	&&+Q_{x_{3} }^{BAB} [1-h(E_{x_{3}}^{BAB} )]- \{Q_{x_{3},n=1 }^{BAE}*h(2e_{1}^{BA} )\nonumber\\
	&&+Q_{x_{3},n=2 }^{BAE}*[\frac{1}{2} h(2e_{2}^{BA} )  +\frac{1}{2}]+Q_{x_{3},n\ge 3 }^{BAE}*1\}.\nonumber
\end{eqnarray}

\begin{figure}[!htbp]\label{diffmu}
	\begin{center}
		\includegraphics[width=9cm,angle=0]{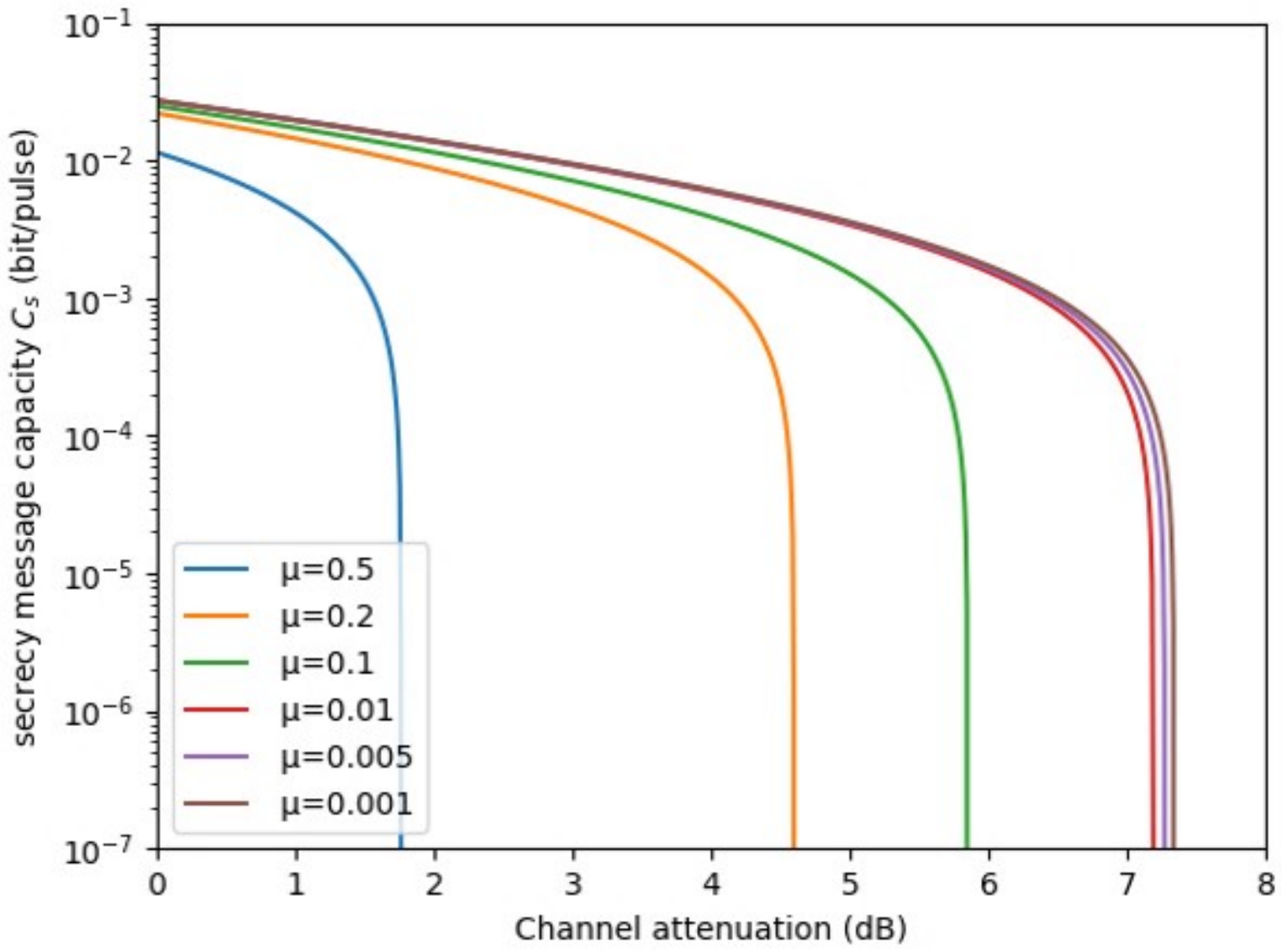}
		\caption{Secrecy message capacity versus the channel attenuation given the collective attack as well as the PNS  attack under the framework of decoy-state analysis. }
	\end{center}
\end{figure}

Fig. 3 illustrates the secrecy message capacity versus the channel attenuation under the collective attack as well as the PNS attack in the framework of decoy-state analysis.
Consistent with traditional GLLP \cite{GLLP} and decoy-state theories \cite{decoy0,decoy}, the maximum communication distances (channel attenuation) of our protocols decrease with the growth of the average photon number $\mu$, due to the increased susceptibility to PNS attacks. Meanwhile, this multiphoton event caused by a high average photon number can also reduce the secrecy message capacity.
In detail, as shown in Eq. (\ref{Cs}), the secrecy message capacity is composed of the difference between $I(A:B)$ and $I(A:E)$. The adoption of HSPS can largely reduce the proportion of vacuum state and increase the proportion of single-photon state in the signal state laser pulses, which can effectively increase $I(A:B)$.  However, the adoption of HSPS cannot reduce the proportion of the multiphoton component. With the growth of the average photon number, the proportion of the multiphoton component increases significantly, which gives Eve more opportunities to steal information and thus increase $I(A:E)$. As a result, our passive QSDC protocol performs better by using the HSPS with low average photon number, in terms of both secrecy message capacity and maximum communication distance. From Fig. 3, the HSPS  with the average photon number $\mu=0.001$ is optimal for our passive QSDC protocol. 

\begin{figure}[!htbp]\label{diffmu}
	\begin{center}
		\includegraphics[width=9cm,angle=0]{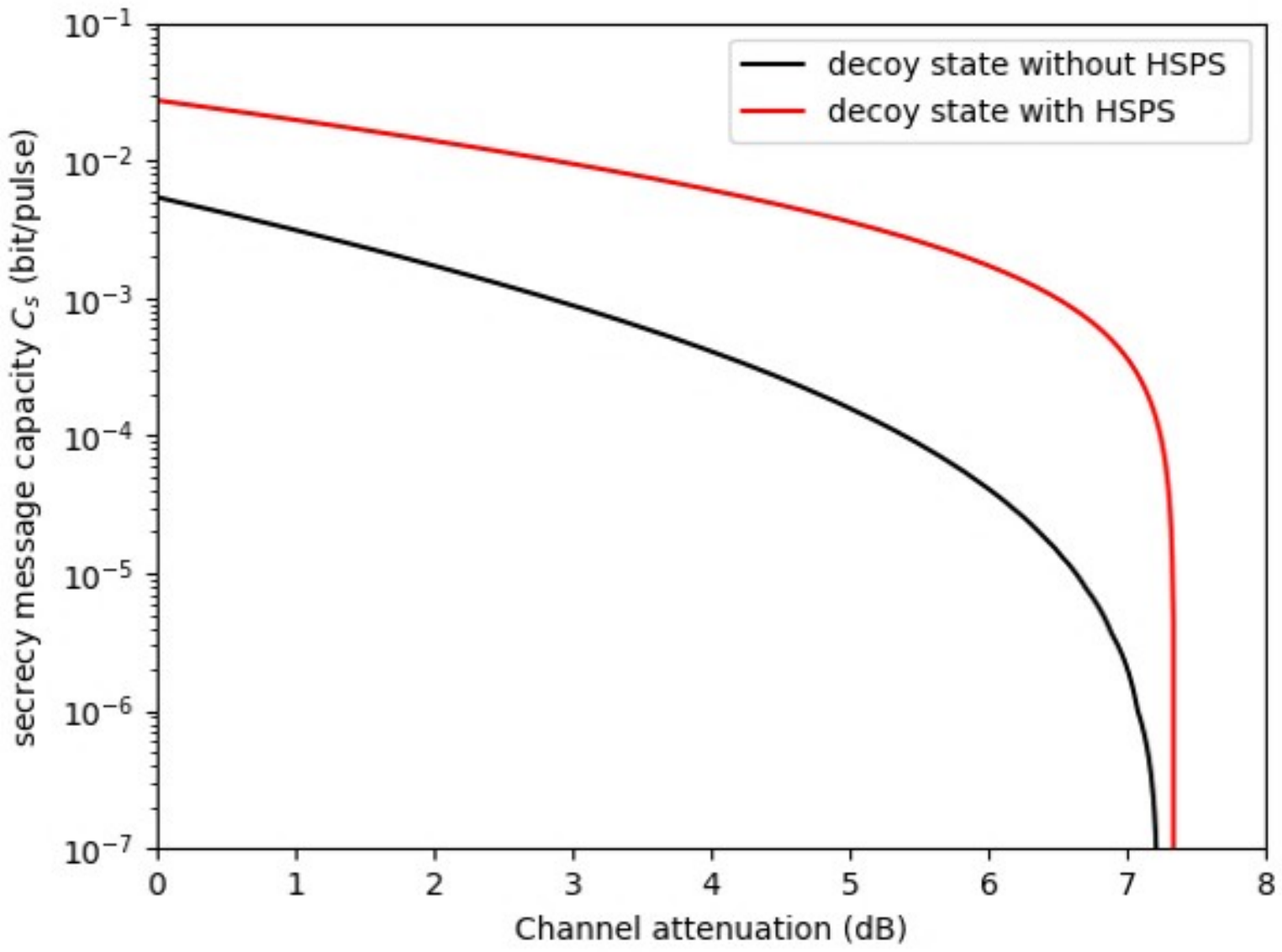}
		\caption{ Comparison of our passive decoy-state QSDC protocol ($\mu=0.001$) with the DL04 QSDC protocol with the WCP source at the optimal intensities. }
	\end{center}
\end{figure}

In Fig. 4, we optimize the DL04 QSDC with the WCP source and decoy-state method (black line) to find the optimal pulse intensity for obtaining the maximal secrecy message capacity at each channel attenuation and compare it with that of our passive decoy-state QSDC protocol with the HSPS  (red line, $\mu=0.001$).
It is evident that the secrecy message capacity of our QSDC protocol is always higher than that of the DL04 QSDC, especially in the case of large channel attenuation. In detail, at the channel attenuation of 4 dB, $C_{s}$ of our QSDC protocol is about 14.92 times greater than that of the DL04 QSDC. At the channel attenuation of 7 dB, $C_{s}$ of our passive QSDC protocol increases to about  186.83  times greater than that of the DL04 QSDC protocol.

Fig. 5 illustrates $C_{s}$ of our passive QSDC protocol with the HSPS and the DL04 QSDC protocol without the HSPS \cite{decoy3} under the average photon numbers $\mu=0.1$ and $0.01$, respectively.
 Comparing with the DL04 QSDC protocol without the HSPS, at a fixed channel attenuation of 4 dB (communication distance of about 10 km), $C_{s}$ of our passive QSDC protocol with the HSPS can be increased to 81.85 times at $\mu=0.1$ and 12.79 times at  $\mu=0.01$. Meanwhile, the maximal communication distance of our passive QSDC protocol with HSPS is also superior to that of the DL04 QSDC protocol without HSPS. When $\mu=0.01$ (low average photon number), the maximal communication distance of our QSDC protocol can achieve  17.975 km (channel attenuation 7.19 dB), which is slightly longer than that of the DL04 QSDC (17.8 km, channel attenuation 7.12 dB). When  $\mu=0.1$ (high average photon number), our  QSDC protocol can achieve the maximal communication distance of about 14.6 km (channel attenuation 5.84 dB), while the DL04 QSDC protocol without the HSPS can only achieve the maximal communication distance of 12.9 km (channel attenuation 5.16 dB).
In Fig. 5, we also compare the secrecy message capacity of our passive QSDC protocol and the DL04 QSDC protocol with infinite decoy state (red lines) and finite decoy state (black lines), respectively. The infinite decoy-state method can lead to the ideal yields and error rates. The simulation results show that our simulated yields and error rates with the finite decoy state are very close to the ideal case with the infinite decoy state. Moreover, under the same average photon-number condition, our passive QSDC protocol is superior to the DL04 QSDC protocol without HSPS with infinite decoy states in terms of the communication distance and secrecy message capacity.

\begin{figure}[!htbp]\label{p2}
	\centering

	\begin{minipage}{\columnwidth}

		\centering
		\includegraphics[width=9cm]{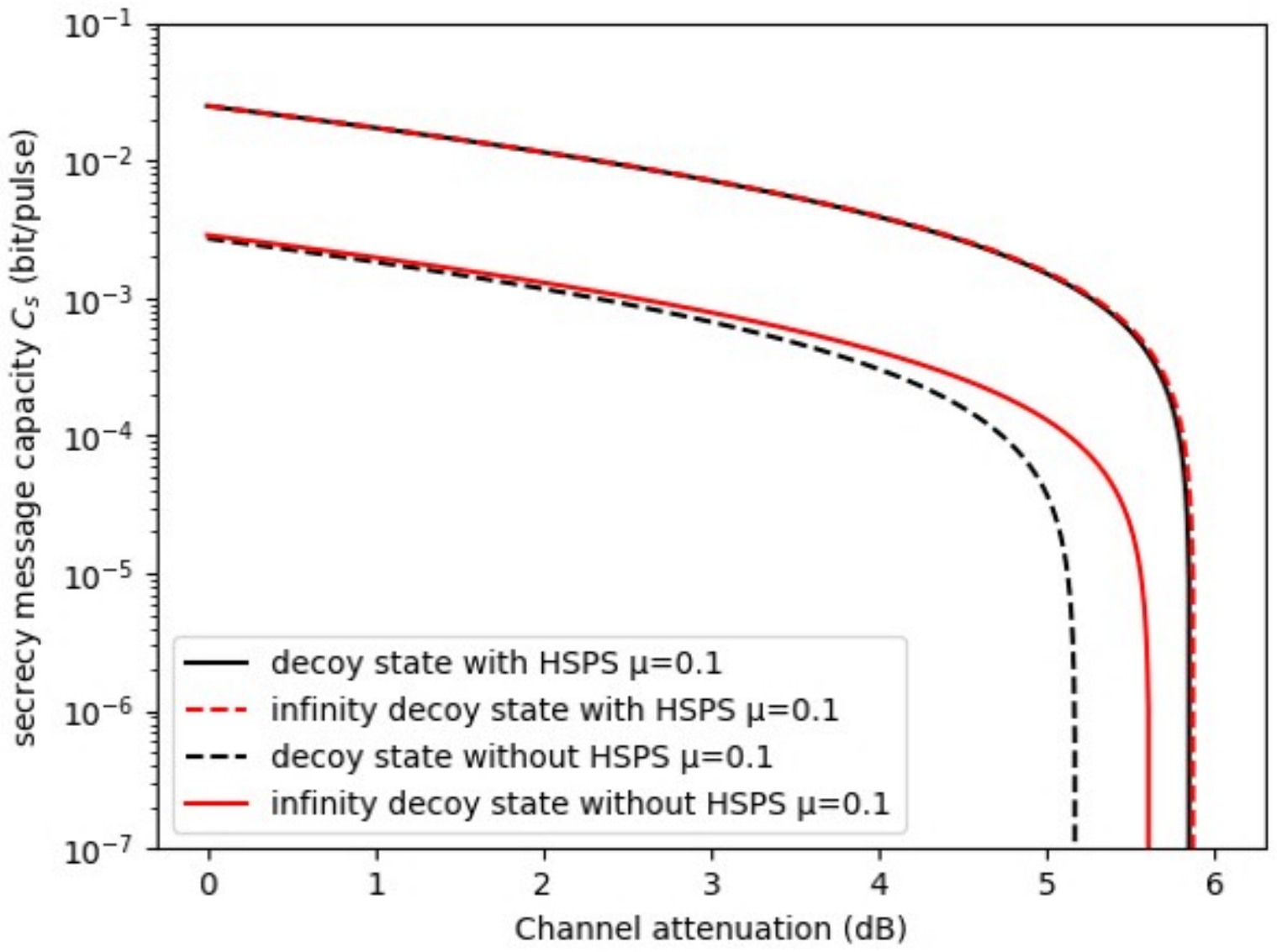}
		\label{fig:image1}
	\end{minipage}
	
	\begin{minipage}{\columnwidth}
		\centering
		\includegraphics[width=9cm]{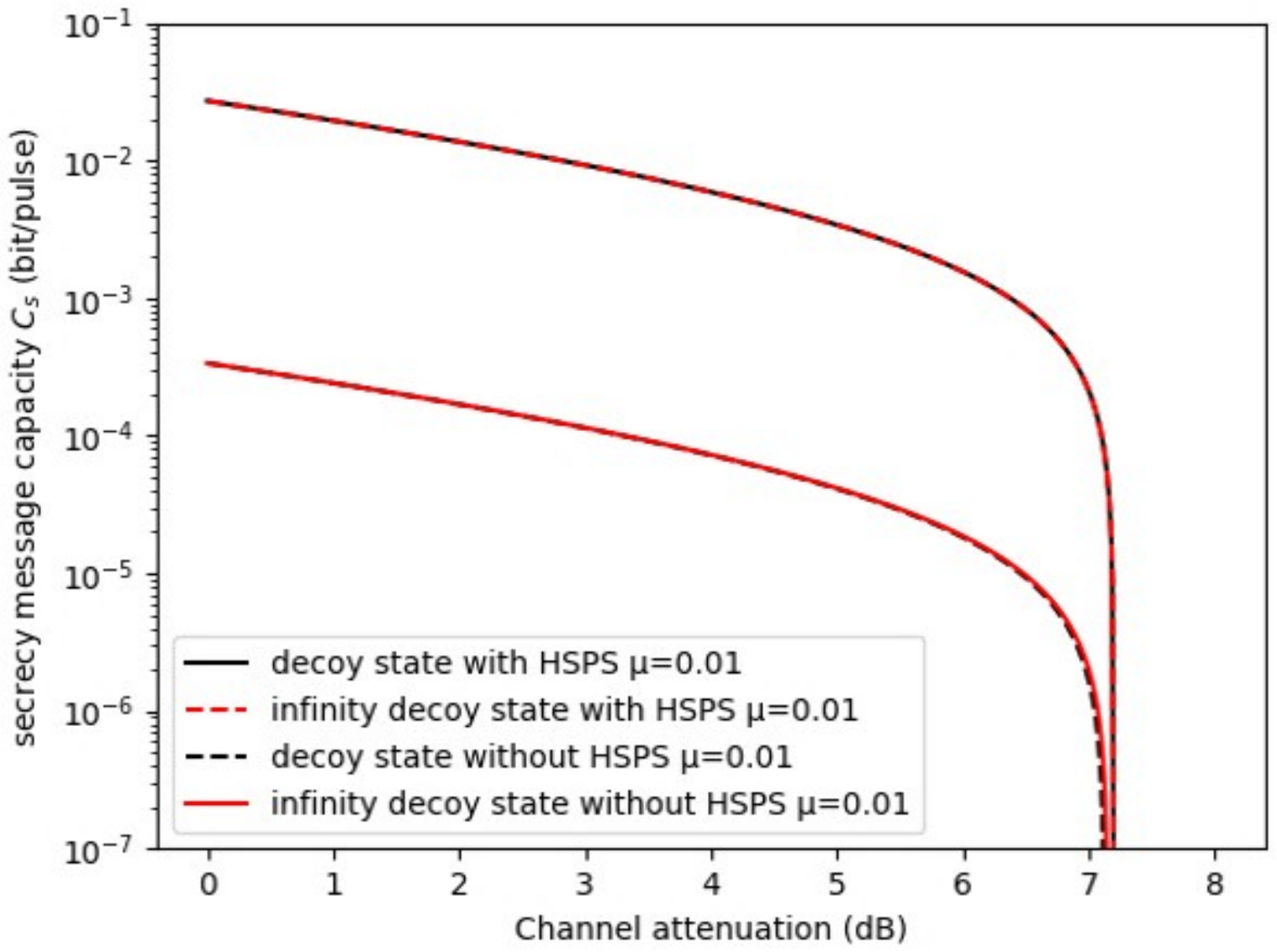}
		\label{fig:image2}
	\end{minipage}
	\caption{ The secrecy message capacity of our passive decoy-state QSDC protocol with the HSPS and the DL04 QSDC protocol without HSPS \cite{decoy3}. The parameters used in our QSDC protocol is also shown in Tab. I. The black and red lines represent the cases of finite and infinite decoy state, respectively.}
	\label{fig:vertical_images}
\end{figure}

\section{Discussion and conclusion}
In QSDC, quantum memory plays a key role for they should ensure that  the quantum channel is secure before transmitting secret messages. Therefore, some photons should be used for security checking and the other photons which are used to encode messages should be stored in quantum memory until the security checking is successful. In previous QSDC experiments, the $^{85}$Rb atoms trapped in a two-dimensional magneto-optical trap (MOT) are used as the quantum memory \cite{QSDC10}. The fiber delay can also act as the role of quantum memory \cite{QSDC9,QSDC11}. 
Moreover, in future multihop quantum communication or quantum network, the quantum repeaters are required. The quantum memory is also indispensable \cite{HSPS5}.
 Quantum memory research has conducted across various physical systems, including atoms \cite{atom,coldatom,coldatom1,qmcell}, defects in solids \cite{qmcen}, hot atomic vapor \cite{qmcell}, superconducting quantum memory \cite{Bao},  and so on.
In practical experiment, the bandwidth mismatch between SPDC source and quantum memory may be an obstacle. Fortunately, many efforts have been  made to couple high-bandwidth SPDC sources with quantum memory \cite{HSPSS1,SPDCS21,SPDCS23,SPDCS24}. 	
For example,  Wei \emph{et al.}  achieved a spectrotemporally  multiplexed quantum memory in cooled erbium-doped silica fiber with bandwidth up to 10 GHz \cite{SPDCS24}.  They utilized dense wavelength division multiplexers to filter out heralded  photons with a bandwidth of 100 GHz.  Then five spectral channels were modulated using optical frequency combs with frequency spacing of 15 GHz. Each channel has a bandwidth of 10 GHz and a separation of 5 GHz. Finally coupled to a multiplexed quantum memory, it can store up to 1650 modes of  heralded single photons.

Moreover,  there have been some efforts to integrate quantum communication with memory, such as memory-enhanced quantum communication \cite{qmqc,qmqkd1,qmqkd2}. The development of these experiments and applications drives the experimental realization of our protocol.
The imperfect quantum memory will affect communication distance and secrecy message capacity of our QSDC protocol. Considering the practical quantum memory in Ref. \cite{HSPSS1}, our protocol can perform about 29.5\% at a channel attenuation of 2 dB (declining from $1.37*10^{-2}$  bit/pulse to $4.05*10^{-3}$  bit/pulse) with the average photon number of $\mu=0.01$. The maximum communication attenuation will also be reduced to 3.82 dB, which is about 53.1\% of that with the perfect quantum memory.
In addition, the quantum-memory-free protocol \cite{QMF1},  coupled with classical cryptography, introduces an alternative approach for implementing our QSDC protocol.  With the classic ciphertext safeguarding the system, even in the absence of quantum memory, Eve can only pilfer the code words, not the meaningful messages, leading to prompt detection of any breach. This approach has been further expanded to entanglement-based two-step protocols \cite{QMF2} and the MDI protocol \cite{QMF3}.
In this way, our passive decoy-state QSDC protocol is hopeful to be demonstrated experimentally in the near future. 

In conclusion, we propose a high-efficient passive decoy-state QSDC protocol with the HSPS. Heralded by the detector responses in the HSPS, the input photon pulse can be passively divided as two kinds of high-quality signal single-photon sources, and a decoy-state source. When neither of the two detectors respond, the input photon pulse in the signal path should be discarded. In this way, the probability of the vacuum state in the signal state and decoy state can be largely reduced. In the security analysis, we consider that Eve performs the PNS
attack combined with the collective attack and calculate the theoretical secrecy message capacity. The simulation results show that our passive decoy-state QSDC protocol with the HSPS is superior to the original DL04 QSDC protocol with WCP source in both secrecy message capacity and maximal communication distance. For a fixed channel attenuation of 4 dB (communication distance of about 10 km), $C_{s}$ of our QSDC protocol can achieve 81.85 times ($\mu=0.1$) and 12.79 times ($\mu=0.01$) of the corresponding values in the original DL04 QSDC protocol with the WCP source. In the high average photon-number condition ($\mu=0.1$), our passive decoy-state QSDC protocol can achieve the maximal communication distance of about 14.6 km, about 1.7 km longer than that of the original DL04 QSDC protocol, while in the low average photon-number condition ($\mu=0.01$), the maximal communication distance can reach 17.975 km.
We also optimize our QSDC protocol and the original DL04 QSDC protocol with the WCP source. Our QSDC protocol is always superior to the DL04 protocol at the optimal intensity. At the channel attenuation of 7 dB, the maximal secrecy message capacity of our QSDC protocol increases to about 186.83 times greater than  that of the DL04 QSDC protocol. Based on the above features, benefiting from the HSPS and passive decoy-state method, our QSDC protocol shows significant advantages in both maximum communication distance and secrecy message capacity and has strong robustness against the side-channel attack. Our passive decoy-state QSDC protocol is conducive to the realization of high-capacity and long-distance QSDC in the future.



\section*{Acknowledgement}
We would like to thank Dong Pan for his generous providing of the comparison data in Fig. 5 and Professors Xiao-Min Hu and Qiang Zhou for helpful discussions. This work is supported by the National Natural Science Foundation of China under Grants  No. 12175106 and 92365110, the Postgraduate Research $\&$  Practice Innovation Program of Jiangsu Province under Grant No. KYCX22-0963, and the Key R$\&$D Program of Guangdong Province under Grant No. 2018B030325002.

\end{document}